\documentclass[superscriptaddress,secnumarabic,amssymb,amsmath,nobibnotes,aps,prd,showkeys,showpacs,nofootinbib,preprint]{revtex4}

\usepackage{graphicx}
\usepackage{float}
\usepackage{bm}
\usepackage{amsmath}
\usepackage{amsfonts}
\usepackage{amssymb}
\usepackage{epstopdf}
\usepackage{natbib}%
\newcommand{\bee}{\begin{equation}}
\newcommand{\eee}{\end{equation}}
\newcommand{\eaa}{\end{eqnarray}}
\newcommand{\baa}{\begin{eqnarray}}

\def\ni{\noindent}
\def\noo{\nonumber \\}

\begin{document}

%\title{\Large Thermostatistical Analysis of Noncommutative Spacetime Schwarzschild Black Holes}
\title{\Large Surface gravity analysis in Gauss-Bonnet \\ and Barrow black holes}
%\title{\Large A note on black holes surface gravity}

\author{Everton M. C. Abreu}\email{evertonabreu@ufrrj.br}
\affiliation{Departamento de F\'{i}sica, Universidade Federal Rural do Rio de Janeiro, 23890-971, Serop\'edica, RJ, Brazil}
\affiliation{Departamento de F\'{i}sica, Universidade Federal de Juiz de Fora, 36036-330, Juiz de Fora, MG, Brazil}
\affiliation{Programa de P\'os-Gradua\c{c}\~ao Interdisciplinar em F\'isica Aplicada, Instituto de F\'{i}sica, Universidade Federal do Rio de Janeiro, 21941-972, Rio de Janeiro, RJ, Brazil}

%\author{Jorge Ananias Neto}\email{jorge@fisica.ufjf.br}
%\affiliation{Departamento de F\'{i}sica, Universidade Federal de Juiz de Fora, 36036-330, Juiz de Fora, MG, Brazil}

%%%%%%%%%%%%%%%%%%%%%%%%%%%%%%%%%%%%%%%%%%%%%%%%%%%%%%%%%%%%%%%%%%%%%%%%%%%%%%%%%%%%%%%%%%%%

%%%%%%%%%%%%%%%%%%%%%%%%%%%%%%%%%%%%%%%%%%%%%%%%%%%%%%%%%%%%%%%%%%%%%%%%%%%%%%%%%%%%%%%%%%
\begin{abstract}
\ni We have different definitions of the surface gravity (SG) of a horizon since we can say we have distinct classifications of horizons.   The SG has an underlying role in the laws of black hole (BH) thermodynamics, being constant in the event horizon.  The SG also acts in the emission of Hawking radiation being connected to its temperature.  Concerning this last issue,
the quantum features that permeate Hawking radiation provide us a direct indication that a BH has its temperature directly connected to its area and that its entropy is proportional to the horizon area.
In this work we analyzed some aspects of event horizons. Analyzing  how the SG can be classically defined for stationary
BHs together with the radial pressure computation.  So, the SG, through the laws of BH mechanics is connected to the real thermodynamical temperature of a thermal spectrum.
We discussed these subjects in two different BHs scenarios, the five dimensional Gauss-Bonnet one and the recently developed Barrow entropy construction.  We discussed how the quantum fluctuations affect these both quantities. 
\end{abstract}
%%%%%%%%%%%%%%%%%%%%%%%%%%%%%%%%%%%%%%%%%%%%%%%%%%%%%%%%%%%%%%%%%%%%%%%%%%%%%%%%%%%%%%%%%%%%
\date{\today}

\maketitle
%%%%%%%%%%%%%%%%%%%%%%%%%%%%%%%%%%%%%%%%%%%%%%%%%%%%%%%%%%%%%%%%%%%%%%%%%%%%%%%%%%%%%%%%%%%%%%%

\section{introduction}

There is a historically well established connection between black hole (BH) evolvement and thermodynamics, and it has been intensely investigated until now since the works of Bekenstein and Hawking \cite{bekenstein,hawking}.  At the beginning, this issue was constructed using the concepts of purely classical evolution \cite{bch}. The posterior disclosure of the semiclassical Hawking radiation effect reordered BH thermodynamics in a stable theoretical position. For a review the interested reader can see \cite{pad-1}. 

One of the consequences of this investigation is that the surface gravity (SG) that appears in the laws of BH mechanics is connected to the actual thermodynamic temperature of a thermal spectrum.   The existence of the Hawking radiation indicates that the development of an astrophysical BH during its lifetime will, in general, ever be completely static.  Namely, we have two distinct situations.  In one circumstance the BH will be evolving by amassing matter (accretion), from either close matter sources or CMB.   Or it will be evaporating, although slowly, through the emission of Hawking radiation.   Considering very small BHs it is assumed that their development is hugely dynamic with a great radiation flux.   To arrange for a BH whose magnitude is neither growing nor declining, it would demand a fine equilibrium between the volume of matter being amassed by accretion and the quantity of energy being lost through Hawking radiation \cite{ny}.   We can also ask what can we say about SG in both situations.

Most computations in the literature concerning the Hawking radiation effect consider both a quasi-stationary and a quasi-equilibrium expansion.    As a matter of fact, many calculations are based on the data of an exactly static background spacetime.   Besides, they neglect any backreaction from the radiation energy flux.  The thermodynamical analogy between BH and thermal systems implies that we should equate both the classical SG and the Hawking radiation temperature.  The SG can be determined classically by using the geometric properties of the fundamental metric.    This same quantity can be demonstrated as being proportional to the temperature of the quantum fields in the Hawking effect. 

As we mentioned before, the SG is defined classically in terms of the geometric features of the fundamental metric.  Although the same quantity appears as the temperature of quantum fields concerning the Hawking effect. It is determined by considering purely geometric quantities like a Killing field, which means that, in standard terms, it is only defined for precisely static scenarios. It is not even described concerning quasi-static situations, where one hopes that the evolution is a quasi-equilibrium picture, and the Hawking radiation is thermal. In this way, the standard description of the SG is not even true in a spacetime that considers gravitational radiation anywhere in the spacetime.   In fact, the SG has importance and, at the same time, distinct roles in BH mechanics. Concerning the classical BH evolution law and considering the so-called zeroth law, this evolution law keeps true independently of any quantum effects and hence, it is essentially a classical effect. The SG acts as the constant of proportionality between the BH mass change and this change occurs in the area. 
It is important for the definition of this zeroth law that the mass of a BH be well defined either.   So, the definition of SG is closely associated with our choice of quasi-local mass.   On the other hand, the SG acts in the emission of Hawking radiation. A suitable generalization of the SG is expected to act in dynamical Hawking radiation, even in non-equilibrium processes \cite{ny}. 

A well known definition of the SG is in terms of a Killing horizon which, for instance, is the way the SG is computed for a Schwarzschild BH (the interested reader can find a good review of Killing horizons properties in \cite{sb}).  This works well in a stationary scenario but spoils in a complete dynamical picture, where no such Killing horizon exists. One way of knowing how the SG works for BHs that are in progress, either by amassing matter or by emitting Hawking radiation was given in \cite{ny}.   However, concerning modified gravitational models, we can say that horizons can be non-Killing or non-null \cite{clv}.

In stationary spacetimes, as the scenario we will analyze here, the BH event horizon is typically a Killing horizon for a convenient chosen Killing vector. On the other hand, for more general, non-stationary spacetimes event horizons, the SG can be determined concerning the null generators of the horizon. 
To consider these more general cases, we have no Killing vector field to use to fix the normalization of the SG. As the generator of the horizon can be defined only on the event horizon, there is no standard way of fixing this normalization by compelling a condition off the horizon.   Recently, big interest has focused on local definitions of horizons \cite{varios-horizon}.  Since the SG is expected to act in the ruling of the quantity of Hawking radiation, it is interesting to analyze SG definitions.

The Gauss-Bonnet (GB) term in $D(>4)$ dimensions is the lowest order correction of the Lanczos-Lovelock (LL) gravity \cite{lanczos,lovelock}.   In both theories the field equations derived from the respective Lagrangians are quasi linear since the initial value problem keeps well defined.   Concerning the horizon thermodynamics described through the Einstein's gravity, it is still true in LL gravity models \cite{psp-mp}.   The interest in $D>4$ GB theories arises from the fact that in $D=4$ it is a pure divergence term, which does not occur in higher dimensions.   The other point of interest is that several string theoretical models have GB-type elements as corrections, what is accomplished after proper field redefinitions \cite{z-dbd}.   The absence of manifest general covariance is not a big problem since this problem also exists for the Einstein-Hilbert action.

In this work, we analyzed some different structures for the SG.   Besides, the GB one mentioned above, another one  appeared recently in the literature, the Barrow toy model \cite{barrow} for BHs, considering a fractal spacetime geometry caused by the quantum fluctuations 
\cite{nossos-meu,nossos,saridakis-ele,saridakis}.  This fractallity feature is reflected in the $\Delta$-parameter present in the definition of the entropy\footnote{In \cite{odintsov-varios} the authors investigated a general expression for the entropy encompassing several entropic models in the literature.  The analysis of these results together with the point of view of this paper is an ongoing research}.  In this paper we make some considerations and compute the SG as a function of this $\Delta$-exponent and the respective radial pressure.  

The distribution of the subjects obeys the following points: in section 2 we considered a 5D Gauss-Bonnet (GB) BH and we calculated its SG and radial pressure.  In section 3 we analyzed Barrow's BH theory and we computed these both quantities to analyze the role of the geometry on both structures.   The final analysis and remarks are described at the conclusion section.

\bigskip

\section{The surface gravity of Gauss-Bonnet black holes}

As explained above, the GB formulation of gravity is a natural generalization of Einstein theory of  relativity.   Since it is the zero-torsion most general theory of gravity, it heads us to the stable second-order equations of motion in higher dimensions.  It was proposed firstly by C. Lanczos \cite{lanczos} and confirmed latter by D. Lovelock in \cite{lovelock}.   The theory includes Einstein's concept of gravity and therefore, GB gravity is more appropriate than other higher-curvature gravitation theories.   Both GB and Einstein terms are within Lovelock's Lagrangian framework.   GB gravitational formulation is the most direct nontrivial generalization of Einstein theory of gravitation.

Having said that, we know that we have in the literature, strong theoretical evidences that there is a strong connection between thermodynamics and the gravitational horizon dynamics.   Alas, this is still not deeply understood.   On the other hand, we know that the Hawking radiation implies that the evolution of an astrophysical BH during its lifetime will almost never be precisely static \cite{ny} or the BH will accretes matter or it will be evaporating via Hawking radiation.

Let us begin with the case of a 5D GB BH explored in \cite{pad-2}, where the GB Lagrangian $L$ in $D$ dimensions \cite{jm,ms} is given by 
\bee
\label{1}
16\,\pi\,L\,=\,R\,+\,\alpha_{{}_{GB}}\,\Big(R^2\,-\,4\,R_{ab}\,R^{ab}\,+\,R_{abcd}\,R^{abcd} \Big)\,\,,
\eee

\ni where $R$ is the $D$ dimensional Ricci scalar and the Lagrangian in Eq. \eqref{1} can be obtained from superstring theory in the low energy limit, $\alpha_{GB}$ is considered as the inverse string tension and it is positive definite \cite{z-dbd}.   The field equations for the semiclassical action obtained using the Lagrangian in Eq. \eqref{1} is
\bee
\label{zero}
G_{ab} \,+\, \alpha_{{}_{GB}} \,H_{ab}\,=\,8\pi\,T_{ab} 
\eee

\noindent where $G_{ab}\,=\,R_{ab}\,-\,\frac 12 g_{ab} R$ and
\bee
\label{0.1}
H_{ab}\,=\,2\Big(R\,R_{ab}\,-\,2R_{ac}\,R^c_{\;\;\;b}\,-\,2R^{cd}\,R_{acbd}\,+\,R_a^{\;\;\;cdm}R_{bcdm}\Big)\,-\,\frac 12 g_{ab}\,L\,\,,
\eee

\noindent where $L$ is given in Eq. \eqref{1}. The static, spherically symmetric BH solutions in this theory have the form
\bee
\label{2}
ds^2\,=\,-\,f(r)\,dt^2\,+\,\frac{dr^2}{f(r)}\,+\,r^2d\Omega_{D-2}^2\,\,,
\eee

\ni where the last term is relative to the angular part of the metric and
\bee
\label{3}
f(r)\,=\,1\,+\,\frac{r^2}{2\,\alpha}\,\Bigg[1\,-\,\sqrt{1\,+\,\frac{4\,\alpha\,\omega}{r^{D-1}}} \Bigg]\,\,,
\eee

\ni where $\alpha=(D-3)(D-4)\,\alpha_{{}_{GB}}$ and $\omega$ is related to the ADM mass $M$ through the equation
\bee
\label{4}
\omega\,=\,\frac{16\pi}{(D-2) V_{D-2}}\,M\,\,,
\eee

\ni where $V_{{}_{D-2}}$ is the volume of unit $(D-2)$-sphere.   The Hawking temperature $T$ and entropy $S$ for this spacetime are, respectively,
\baa
\label{5}
 T&=& \frac{D-3}{4\pi r_+}\Bigg[\frac{r^2_+}{r_+^2 + 2\alpha}\,+\,\alpha\Bigg(\frac{D-5}{D-3} \Bigg)\,\frac{1}{r^2_+ + 2\alpha} \Bigg]
\eaa
and
\baa
\label{6}
S&=&\frac{{\cal A}}{4}\Bigg[1\,+\,2\alpha\,\Bigg(\frac{D-2}{D-4} \Bigg)\,\Bigg(\frac{{\cal A}}{\Sigma_{{}_{D-2}}} \Bigg)^{-\frac{2}{D-2}} \Bigg]\,\,,
\eaa

\ni where ${\cal A}\,=\, V_{{}_{D-2}}\,r_+^{D-2}$ is the horizon area.  We can also realize that $D=4$ is a divergent point in GB gravity. 

Several authors dedicated to calculate the entropy of GB theory \cite{entropyGB}.  Therefore we can see, as well known, that in such general theories, different from Einstein's gravity, the horizon entropy is not proportional to the area, it only occurs when $\alpha=0$ in Eq. \eqref{6} \cite{pad-2}. The locus of the horizon is found as roots of $q(r_+)=0$, where
\bee
\label{7}
q(r)\,=\,R^{D-3}\,+\,\alpha\,r^{D-5}\,-\,\omega
\eee

\ni and for the horizon to exist at all, one must also have that
\bee
\label{8}
r_+^2\,+\,2\,\alpha \geq 0\,\,.
\eee

In this generalized GB BH, let us compute the SG
\bee
\label{9}
\kappa\,=\,\frac 12 \frac{df(r)}{dr}\Bigg|_{r=r_+} \,=\, \frac 12 f'(r)\Big|_{r=r_+}  \,\,,
\eee

\ni where the derivative of $f(r)$ is given by
\baa
\label{10}
&&f(r)\,=\,1\,+\,\frac{r^2}{2\alpha}\,\Big[1\,-\,\sqrt{1\,+\,Ar^{1-D}}\,\Big] \noo
&&\mbox{} \noo
&\Longrightarrow& f'(r)\,=\,\frac{r}{\alpha}\Bigg\{1\,-\,\sqrt{1\,+\,Ar^{1-D}}\,+\,\frac r2\Bigg[\frac A2 (1-D)r^{-D}\Big(1\,+\,Ar^{1-D}\Big)^{-\frac 12} \Bigg]\Bigg\}\,\,.
\eaa

\ni where $A=4\alpha\,\omega$.   We can see in Eq. \eqref{10} the SG in Eq. \eqref{9} is directly dependent of the dimensionality of the geometry.   Besides, it is a function of the ADM mass via the $A$ term.   We can also see that for $D=3$ or $D=4$ we have an infinite value for SG since $\alpha \rightarrow 0$, which means form Eq. \eqref{8} that $r_+ \geq 0$, which, from Eq. \eqref{5}, results in a plausible value for the Hawking temperature and from Eq. \eqref{6}, the Bekenstein-Hawking entropy is recovered.   For $r_+ \rightarrow 0$, i.e., very small BHs, for a $D\not= 3$ GB BH, from Eq. \eqref{5} we confirm the result that BHs are very hot at the event horizon.   Since the entropy is not dependent directly of $r_+$, it keeps the Bekenstein-Hawking form when $D\not= 3$ and $\alpha \rightarrow 0$.

The Einstein equation for this metric is
\bee
\label{A-1}
rf'(r)\,-\,\Big(1\,-\,f(r)\Big) \,=\,8\pi\,P\,r^2 \,\,,
\eee

\ni where $P=P(r)$ is the radial pressure.   But we know that $f(r)$ is zero at the horizon, i.e., in $r=r_+$ and we have that $f(r_+)\,=\,0$, hence, using this information in Eq. \eqref{A-1} we have that
\bee
\label{A-2}
r_+\,f'(r_+)\,-\,1 \,=\, 8\pi P\, r_+^2\,\,,
\eee

\ni where we are using that $G=c=\hbar=1$.  Hence, from the above equations we can write the radial pressure as
\bee
\label{A-3}
P(r_+)\,=\,\frac{1}{8\pi r_+^2}\,\Big(r_+ f'(r_+)\,-\,1\Big)\,\,,
\eee

\ni and from Eq. \eqref{9} the final expression for the SG is given by
\bee
\label{A-4}
\kappa\,=\,\frac 12 f'(r_+)\,=\,  \frac{r_+}{2\alpha}\Bigg\{1\,-\,\sqrt{1\,+\,Ar^{1-D}_+}\,+\,\frac{r_+}{2}\Bigg[\frac A2 (1-D)r^{-D}_+\Big(1\,+\,Ar^{1-D}_+\Big)^{-\frac 12} \Bigg]\Bigg\}\,\,\,\,.
\eee

\noindent The zeroth law of BH thermodynamics asserts the $\kappa$ is constant in the event horizon, which means that we have two alternatives.   Firstly all the terms in Eq. \eqref{A-4} are constant, obviously.   The other one is that there is a compensation such that the final result keeps constant.   Let us consider the first alternative as the correct one,   
So, $\omega$ in Eq. \eqref{4} is also constant which means that the mass $M$ is constant and all the terms in Eq. \eqref{A-4} are constants.  Hence, since the horizon radius $r_+$ is constant, from Eq. \eqref{A-3} we can see that the radial pressure is also constant in the event horizon.

\section{Barrow black hole}

Barrow introduced recently in 2020 \cite{barrow}, based on the Covid-19 shape of the virus, which shows a fractal structure, i.e., a fractal framework for the horizon of BHs with finite volume and infinite, or not, area.  As a toy model, it can show effects of quantum gravity spacetime foam, with important consequences concerning the entropy of BHs and the Universe.   The so-called Barrow-parameter, as will be shown just below, is confined within $0\leq \Delta \leq 1$, where $\Delta=0$, geometrically speaking, responds for a smooth spacetime structure and $\Delta=1$ shows a most intricate geometry.   As demonstrated in \cite{barrow}, the Hawking lifetime of BHs is also shortened, but we do not discuss this issue here.

Now, let us analyze the Barrow BH with fractal geometry 
\bee
\label{8}
S\,=\,\Bigg(\frac{A_{{}_H}}{4A_{{}_{Pl}}}\Bigg)^{1+\frac \Delta2}\,\,,
\eee

\ni where $A_{{}_{Pl}}=\ell^2_P$ is the Planck area, which is equal to $1$ in relativistic dimensions.
%$A_{Pl}=4\,\ell_{P}ˆ{2}$.  
Thermodynamically speaking, for $\Delta=0$ we have the Bekenstein-Hawking entropy and for $\Delta=1$, the entropy for the most intricate geometry.   For $A_{{}_H}=4\pi r_+^2$ we can write the entropy derivative in a convenient form such that
\bee
\label{A0}
dS\,=\,2\,\Bigg(1\,+\,\frac \Delta2 \Bigg)\pi^{1+\frac \Delta2}\,r_+^\Delta \,(r_+\,dr_+)\,\,,
\eee

\ni and the Einstein equations, with an ``Ad hoc" $dr_+$-term in each factor of the equation,
\bee
\label{A}
\frac 12 f'(r_+)\,r_+dr_+\,-\,\frac 12 dr_+\,=\,4\pi P \,r^2_+ \,dr_+\,\,,
\eee

\ni and we will use this equation including the radial pressure in a moment.  We know that the periodicity in Euclidean time permits us to connect the horizon to the temperature such that
\bee
\label{B}
k_{{}_B}\,T\,=\,\frac{\kappa}{2\pi}\,=\,\frac{f'(r_+)}{4\pi}\,\,.
\eee

Let us use the first law equation
\bee
\label{C}
k_{{}_B} T\, dS\,-\,dE\,=\,P\,dV\,\,,
\eee

\ni which is analogous to Eq. \eqref{A} \cite{pad-3} and, using also Eqs. \eqref{A0}-\eqref{C} we have that
\bee
\label{C-1}
f'(r_+)\,=\,2\kappa\,\Big(1\,+\,\frac \Delta2\,\Big)\pi^{\Delta/2}\,r_+^\Delta
\eee

\ni and hence, the SG for Barrow's BH is
\bee
\label{D}
\kappa_{{}_{Barrow}}\,=\,\frac{f'(r_+)\,r_+^{-\Delta}}{2\Big(1+\frac \Delta2 \Big) \pi^{\Delta/2}}\,\,.
\eee

\ni  Let us now analyze the geometrical features coming from Barrow structure.
For $$\Delta=0 \qquad \Longrightarrow \qquad \kappa_{{}_{Barrow}}\,=\,\kappa\,=\,\frac{1}{2}\,f'(r_+)\,\,,$$ as expected.   For $\Delta=1$ we have that
\bee
\label{E}
\kappa_{{}_{Barrow}}\,=\,\frac{1}{3\sqrt{\pi}}\frac{f'(r_+)}{r_+}
\eee

\ni where the presence of $r_+$ shows that in a fractal spacetime the SG expression is dependent of the radius, what does not happen in the flat spacetime.

From Eq. \eqref{B} the temperature of the horizon is
\bee
\label{F}
T_{{}_{Barrow}}\,=\,\frac{1}{2\pi k_{{}_B}}\,\kappa_{{}_{Barrow}}\,\,,
\eee

\ni where $T_{{}_{Barrow}}$ is also an effective temperature.   It differs by the same factor from the thermodynamical temperature
\bee
\label{G}
t_{{}_{Barrow}}\,=\frac{T_{{}_{Barrow}}}{\sqrt{1\,-\,\frac{2E}{r}}}\,\,,
\eee

\ni measured by constant-radius detectors \cite{bd}.

It is well known that the SG is also constant for stationary BH and, in this way, linking both zero laws.   in this case, from Eq. \eqref{D}
\bee
\label{H}
\frac{df}{dr}\Big|_{r=r_+}\,=\,2\,\kappa_{{}_{Barrow}}\Big(1\,+\,\frac \Delta2 \Big)\,\pi^{\Delta/2}\,r^\Delta \Big|_{r=r_+}\,\,,
\eee

\ni which allows us to write a general relation
\bee
\label{I}
\frac{df(r_+)}{dr}\,=\,2\,\kappa_{{}_{Barrow}}\Big(1\,+\,\frac \Delta2 \Big)\,\pi^{\Delta/2}\,r^\Delta_+\,\,,
\eee
\bee
\label{J}
\Longrightarrow \qquad f(r_+)\,=\,\frac{2}{\Delta+1}\,\kappa_{{}_{Barrow}}\Big(1\,+\,\frac \Delta2 \Big)\,\pi^{\Delta/2}\,r^{\Delta+1}_+\,+\,\Lambda_o\,\,,
\eee

\ni where $\Lambda_o$ is a constant that depends on the initial conditions of $f(r)$.

Hence, from Eqs. \eqref{A-1} and \eqref{I} we can calculate the radial pressure
\bee
\label{K}
P(r)\,=\,\frac{1}{8\pi}\,\frac{\Delta+2}{\Delta+1}\,\frac{1}{r^2}\Bigg(r^{\Delta+1}\,-\,\frac{\Delta+1}{\Delta+2}\Bigg)\,\,,
\eee

\ni as a function of the radius and the $\Delta$-parameter.  We can see clearly that the radial pressure depends directly on the structure of the spacetime represented by the $\Delta$-parameter in Barrows's formulation.  Moreover we can see that for $r \rightarrow \infty$ the radial pressure goes to zero which is coherent to the fact that as the radius strongly grows, the radial pressure diminishes.

\section{Conclusions, final remarks and perspectives}

In this work our objective was to show precisely that using the recently developed Barrow's formulation of BHs, we can see that some aspects of the event horizon are affected by the quantum fluctuations that dictate the geometry.  In some papers published by us before \cite{nossos}, the thermodynamics of BHs were discussed, but some issues relative exactly to the event horizon was now computed.

We have analyzed the SG issue concerning two popular BH formalisms, the GB and the Barrow toy model, which shows that the geometry affects directly the entropy of the BH.   Our main motivation concerning SG is because it is a relevant subject since it is equivalent to the thermodynamic temperature in the investigation of BH thermodynamics laws, more specific, the zeroth-law where a direct association between SG and the thermodynamic temperature appears.

We calculated the expressions of the SG and radial pressure for both models.  We saw that both are dependent of the dimensionality and the structure of the spacetime, in the Barrow's case.   In GB case, we saw that the SG remains constant since all the terms are constant or if there is a compensation procedure concerning accretion and radiation processes, and consequently the zeroth-law can be obeyed.   By analyzing the final expressions of SG, using particular values for some parameters connected to the dimensionality and the ADM mass,   The results recover the well known results that shows that the event horizon is a very hot geometrical place.

In Barrow's case, the final expressions of both SG and radial pressure show a strong dependence of geometrical structure through the presence of $\Delta$ since we have to consider $\Delta$ as a constant term because of the zeroth-law.  So, an analysis like the one made in \cite{nossos-meu,saridakis-ele} is not possible directly, but maybe a compensating factor relative to accretion and radiation simultaneous effects can be constructed.

Another perspective is relative to how can the expressions obtained here affects the other laws of BH thermodynamics since we have modified the SG as well as the entropy.  The thermodynamical potentials can also be studied.   Other gravitation models can also be analyzed, of course.   On the other hand, we can also include numerical values in our results and obtain a value for $\Delta$, for example.

\end{document}